\documentclass{PoS}
\pdfoutput=1
\usepackage{epsfig}
\usepackage{graphicx}

\usepackage{url} 
\usepackage{wrapfig}
\usepackage{subfigure}
\usepackage{amsmath}
\usepackage{amssymb}
\usepackage{appendix}
\usepackage{alltt}
\usepackage{longtable}
\usepackage{tabularx}

\ifpdf\else\usepackage{graphics}\fi

\makeatletter
\renewcommand{\fnum@figure}{{\bf Figure \thefigure}}
\renewcommand{\fnum@table}{{\bf Table \thetable}}
\makeatother
\let\ocaption\caption
\renewcommand{\caption}[2][]{\ocaption[#1]{{\small\it #2}}}
\title{Statistical inversion of the LOFAR Epoch of Reionization experiment data model}

\ShortTitle{Statistical inversion of the LOFAR EoR data}

\author{\speaker{Panos Lampropoulos}\\
ASTRON, Oude Hoogeveensedijk 4, 7991 PD, Dwingeloo, the Netherlands\\
 Kapteyn Astronomical Institute, Landleven 12, 9747 AD, Groningen, the Netherlands\\
                E-mail: \email{labropoulos@astron.nl}}

\author{the LOFAR EoR Key Science Project team\\}
\abstract{LOFAR (\textbf{LO}w \textbf{F}requency \textbf{AR}ray) is a new and innovative effort to build a radio-telescope operating at the multi-meter wavelength spectral window. One of the most exciting applications of LOFAR will be the search for redshifted 21-cm line emission from the Epoch of Reionization (EoR). It is currently believed that the Dark Ages, the period after recombination when the Universe turned neutral, lasted until around the Universe was 400,000 years old. During the EoR, objects started to form in the early universe and they were energetic enough to ionize neutral hydrogen. The precision and accuracy required to achieve this scientific goal, can  be essentially translated into accumulating large amounts of data.  The data model describing the response of the LOFAR telescope to the intensity distribution of the sky is characterized by the non-linearity of the parameters and the large level of noise compared to the desired cosmological signal.   In this poster, we present the implementation of a statistically optimal map-making process and its properties. The basic assumptions of this method are that the noise is Gaussian and independent between the stations and frequency channels and that the dynamic range of the data can been enhanced significantly during the off-line LOFAR processing. These assumptions match our expectations for the LOFAR Epoch of Reionization Experiment. }

\FullConference{ISKAF2010 Science Meeting - ISKAF2010\\
		June 10-14, 2010\\
		Assen, the Netherlands}

\begin{document}
\label{firstpage}
\maketitle

\section{Introduction}
The Epoch of Reionization (hereafter EoR)  demarcates the phase transition of the Universe, during which the hydrogen gas turned from the neutral to the ionized state. The last two decades have seen an increased theoretical effort to understand the interplay between the physical processes that characterize the EoR. Nonetheless, observational evidence is still scarce and indirect \cite{furla}. The LOw Frequency ARray (LOFAR) is one of the currently designed instruments with the aim to probe the end of the Dark Ages and cosmic reionization and will reach the final stages of its construction at the end of this year (2010).

One of the most challenging aspects of the LOFAR EoR experiment is the large dynamic range between the different components of the sky signal. Discrete sources can be of the order of $10^{1-5}\mathrm{Jy/beam}$ while the Galactic diffuse emission, as well as the confusion,  amount to $5 \mathrm{mJy/beam}$ and $3 \mathrm{mJy/beam}$ respectively\footnote{We have assumed a PSF of three arcminutes, so that   $1 \mu \mathrm{Jy/beam}$ corresponds to 2 Kelvin.}. The noise in the data is of the order  $10{\mu \mathrm{Jy/beam}}$, while the desired cosmic signal of the order of $1 \mu {\mathrm{Jy/beam}}$, where we  assumed a synthesized beam resolution of 3 arcminutes at 150 MHz. Even after very accurate foreground removal the EoR signal is still buried deep in the noise. To reach the sensitivity required to statistically detect the EoR signal, a long observation run of  at least 400 hours is necessary. This will result in a recorded visibility dataset of the order of one to two petabytes, including calibration and flagging meta-data.

One might therefore ask the question: What are the prospects and limitations of EoR  imaging or power-spectrum estimation with a phased array? Even if the array is perfectly calibrated the result will be noise-limited at diffraction-limited resolution. The fidelity of the map is difficult to assess even in this simple case, due to the improper sampling of the uv plane. This leads to a number of questions regarding the calibration:
\begin{itemize}
 \item How can we correct, to the desired level of accuracy, for the systematic errors in the presence of noise?
\item How  can we achieve this in a computationally efficient way, given the large number of visibilities that the experiment produces.
\item What is the theoretical limit on calibration accuracy (expressed via the information theoretic Cram\'er-Rao bound) and can we reach it in practice? 
\item Is the detection of the EoR signal feasible under the various instrumental distortions, incomplete uv-plane sampling and large noise power?
\end{itemize}

\section{Maximum Likelihood inversion}
In array signal processing, the problem of estimating the parameters of multiple sources emitting signals that are received  is addressed. There are several estimators proposed in the literature, but in this work we focus on the Maximum Likelihood (ML) estimator. The method was pioneered by R. A. Fisher in 1922.  The maximum likelihood estimator (MLE) selects 
the parameter value which gives the observed data the largest possible probability density in the absence of  prior information, 
although the latter can be easily incorporated to make the analysis fully Bayesian. For small numbers of samples, the bias of maximum likelihood 
estimators can be substantial, but for fairly weak regularity conditions it can be considered asymptotically optimal.  Thus, for large samples the MLE has been shown to achieve the Cram\'er-Rao lower bound and yield asymptotically efficient parameter estimates \cite{stoica}. With large numbers of data points (of the order of $10^9$ contrasted to $10^6$ estimated parameters), as in the case of the LOFAR EoR KSP,  the bias of the 
method tends to be very small (Central Limit Theorem).

This immense amount of data is affected by instrumental corruptions, which will be determined, to first order, during the initial processing. This involves finding a good initial solution of the parameters for all instrument
and sky effects using a modified SELFCAL loop
and a simple model for for the sky (e.g. bright calibrator sources). Solving for the parameters is a highly non-linear process, bound to converge to secondary minima, if not carried out carefully.

The data model can be written as a set of linear equations $\mathbf{v}=\mathbf{A}\left( {\bf{p}} \right)\mathbf{s}+\mathbf{n}$, where $\mathbf{v}$ is the observed data vector, $\mathbf{A}\left( {\bf{p}} \right)$ is a sparse matrix describing the instrumental effects, $\mathbf{s}$ is the true underlying sky signal and $\mathbf{n}$ is a vector representing uncorrelated, spatially white noise. The Maximum Likelihood solution to this problem is \cite{panos}:
\[{\bf s}_{\rm ML} = \left[ {{\bf{A}}^{\dag} \left( {\bf{p}} \right){\bf{C}^{-1}_{\rm noise}} {\bf{A}}\left( {\bf{p}} \right)} \right]^{-1}
\left[ {{\bf{A}}^{\dag} \left( {\bf{p}} \right){\bf{C}^{-1}_{\rm noise}} } \right]{\bf{v}}.\] 
 $\bf{C}^{-1}_{\rm noise}$ stands for the inverse covariance matrix of the noise. Solving this equation is essentially a linear algebra problem, but the solution is non-trivial because  $\bf{v}$ is a vector of $10^9$ double-precision, complex numbers.

\section{Regularization}
The system of linear equations that we described in the previous section cannot be inverted directly. The resulting system matrix is singular for most practical cases and thus regularization has to be used in order to get an approximate solution that is close to the real one. 
The regularized Maximum Likelihood solution becomes:
\[
{\mathbf{A}}^T {\mathbf{C}}_{{\text{noise}}}^{ - {\mathbf{1}}} {\mathbf{v}} = \left( {{\mathbf{A}}^T {\mathbf{C}}_{{\text{noise}}}^{ - {\mathbf{1}}} {\mathbf{A}} + \alpha {\mathbf{R}}} \right){\mathbf{s}},
\]
where $\alpha$ is a Lagrange multiplier (regularization parameter) and ${\mathbf{R}}$ is the regularization functional. For the regularization functional we use two methods: Tikhonov and diffusion operators that take into account the local curvature of the data and the measurement equation.
\begin{figure}

\centering
\includegraphics[scale=0.5]{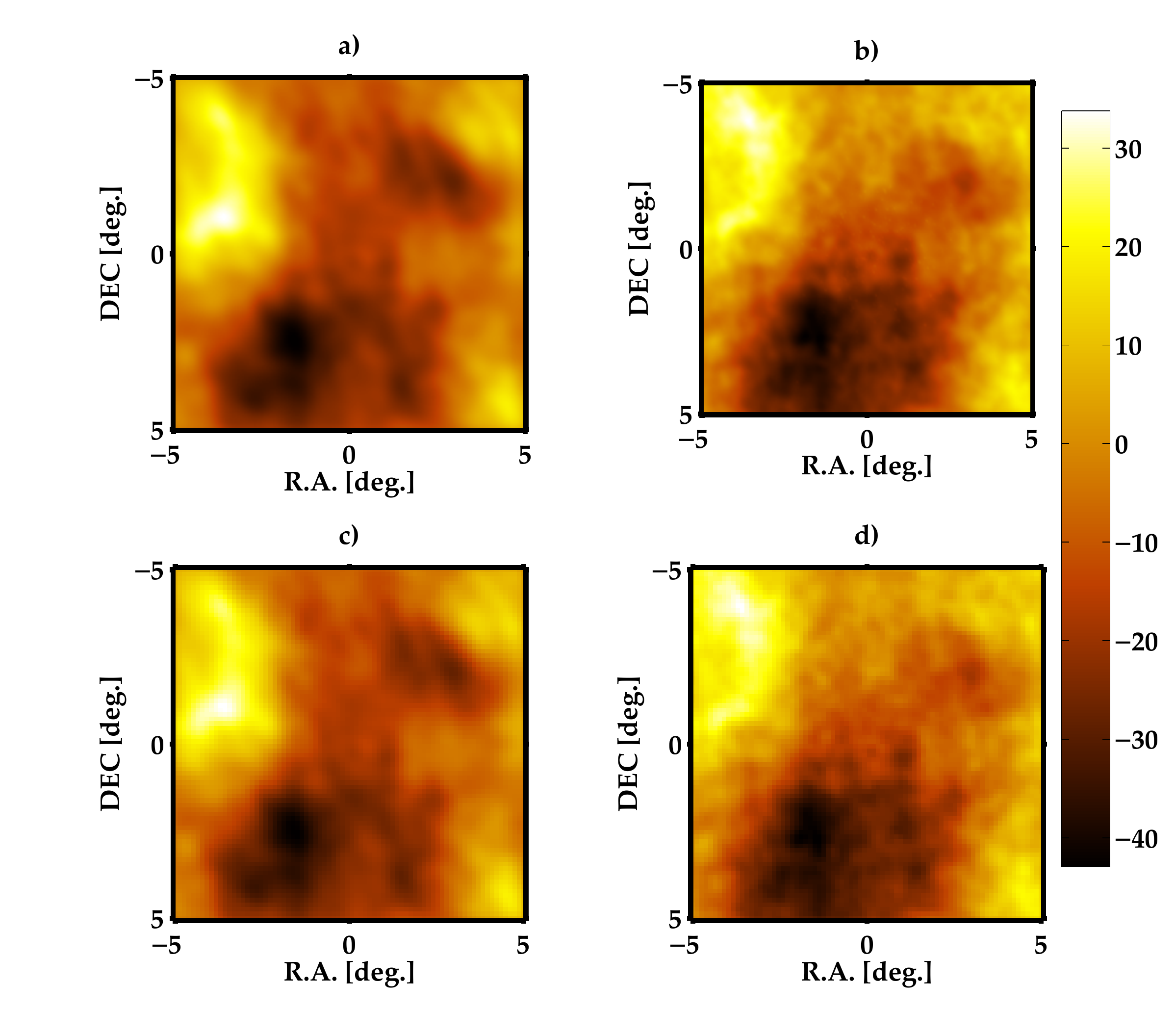}

\caption{Foreground maps generated using different regularization methods and resolutions. The top row  shows the reconstruction using diffusion operators and the bottom row using Tikhonov functionals. The maps at the left column are at a resolution of $80\times80$ pixels and on the right at $160\times160$ pixels: }
\label{reg1}
\end{figure}

Figure \ref{reg1} shows the comparison of the Tikhonov and diffusion regularization methods for final maps for different map resolutions. We see that Tikhonov regularization is particularly harsh on suppressing extreme values. The MLE is performed on the uv-plane and we present the direct Fourier transform of the visibilities (dirty map). To get statistically optimal  results the dynamic range of the data has to be 100:1 or larger. 

\section{Cram\'er-Rao lower bound and the power-spectrum}
In order to demonstrate the potential gain in estimation accuracy of the ML estimator we need to evaluate numerically the asymptotic covariance. In order to calculate the Cram\'er-Rao bound (CRB), which states that the variance of an unbiased estimator $\hat\vartheta$ is bounded by the inverse of the Fisher information matrix (FIM)  $F(\vartheta)$, the FIM has to be computed. The elements of the inverse CRB that is the FIM are given by the Bangs formula \cite{stef}:
\[
\begin{gathered}
  \mathbf{F}  = \left( {\frac{{\partial \mathrm{vec}{\mathbf{\hat V}}_{obs} }}
{{\partial {\mathbf{m}}}}} \right)^T \left( {{\mathbf{\hat V}}_{obs}^{ - T}  \otimes {\mathbf{\hat V}}_{obs} } \right)\left( {\frac{{\partial \mathrm{vec}{\mathbf{\hat V}}_{obs} }}
{{\partial {\mathbf{m}}}}} \right) \hfill \\
  , {\mathbf{m}} = \left[ {\begin{array}{*{20}c}
   {\mathbf{g}} & \theta  & \sigma,   \\

 \end{array} } \right], \hfill \\ 
\end{gathered} 
\]
where $\mathbf{m}$ is the instrumental parameter vector and $g,\theta$ and $\sigma$ describe uv-plane, image-plane effects and noise. The CRB is obtained by applying block-matrix inversion identities.

\begin{figure}
\centering

\centering
\includegraphics[scale=0.5]{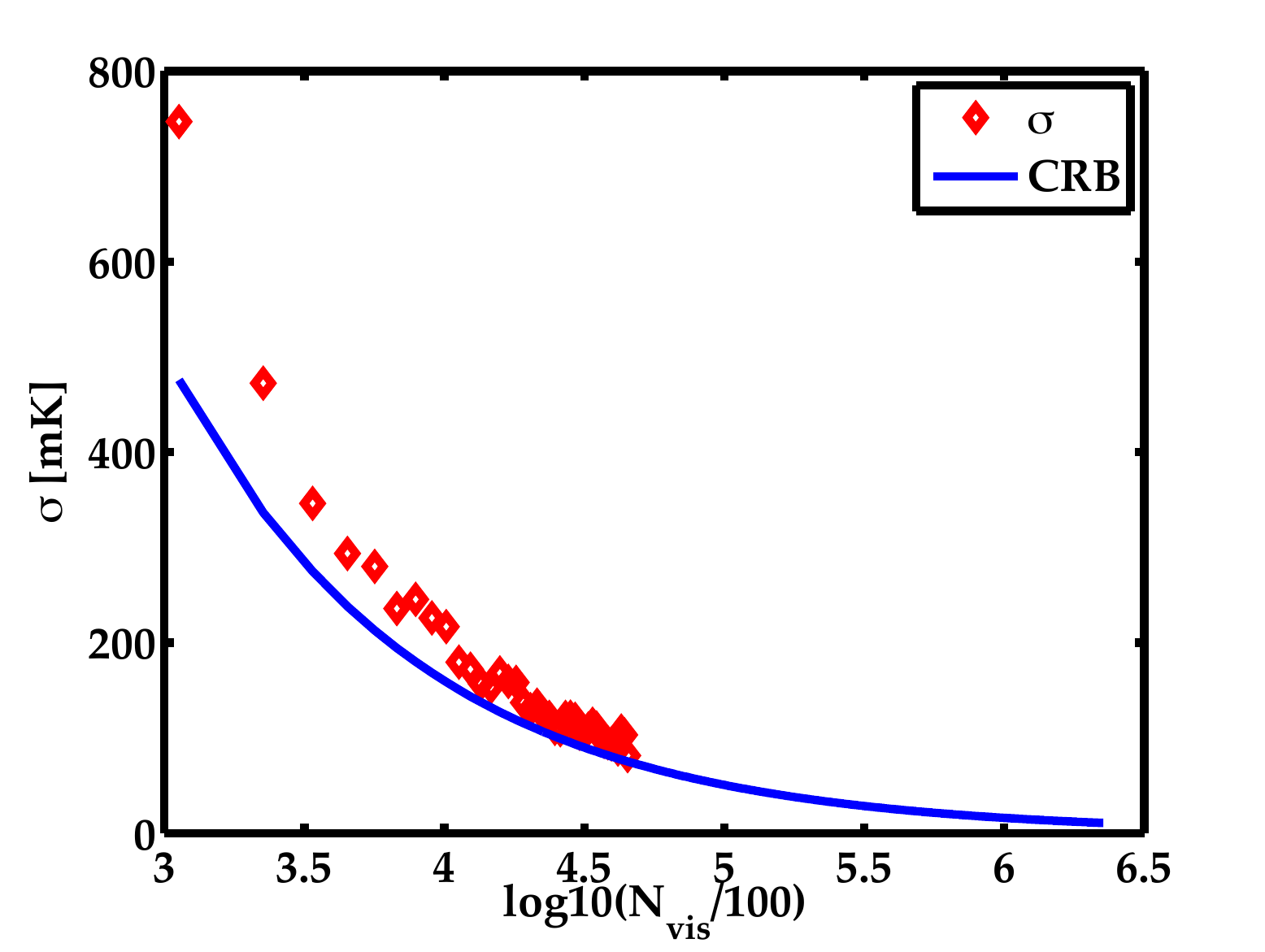}

\caption{Bounds on the standard deviation of the signal power. Thirty seconds of averaging and 0.5 MHz of bandwidth are assumed.  }
\label{crb}
\end{figure}

In Figure \ref{crb} we present the variance of the noise and the CRB as a function of SNR. In the case of interferometry the SNR scales with the number of data points as $N^{-\frac{1}{2}}$. Thus we can move to the asymptotic convergence regime by either integrating more or increasing the number of stations.We have assumed that we observe for 4 hours per night using 0.5 MHz of bandwidth and 30 seconds of averaging. This means that within each integration time interval we accumulate $\sim 10^4$ visibilities per frequency. The blue line shows the CRB computed from the parameters obtained from a realistic EoR simulation described in \cite{panos}. We then estimate the standard deviation using a varying number of visibilities to generate maps. We see that we approach the CRB for a number of visibilities that approaches $\sim 10^9$, which corresponds to 400 hours of integration. However, we must raise attention to the following issues: In this work we have ignored RFI and mutual coupling. These effects can add coherent contamination to the observed visibility and that would result in a Fisher information matrix with less than full-rank. In that case the MLE estimator would not be unbiased, although it might exhibit the same asymptotic behavior. The effect is more prominent for dipoles and tile correlation that are relatively close to each other, but for stations that are separated by a large enough distance this should not be a problem.  This issue must be addressed through new, more sophisticated simulations. A Monte-Carlo type simulation is also needed in order to calculate the global FIM.

\subsection{EoR signal extraction}
The ultimate benchmark of the EoR experiment is set by the ability to extract the cosmological HI signal. In this section we perform a simple extraction of the signal using a polynomial fitting method in the uv-plane \cite{harker09b}. To do that we need to select scales on the uv-plane that are sampled for a large fraction of the available observation bandwidth and with high SNR.
 
\begin{figure}
\centering

\includegraphics[scale=0.5]{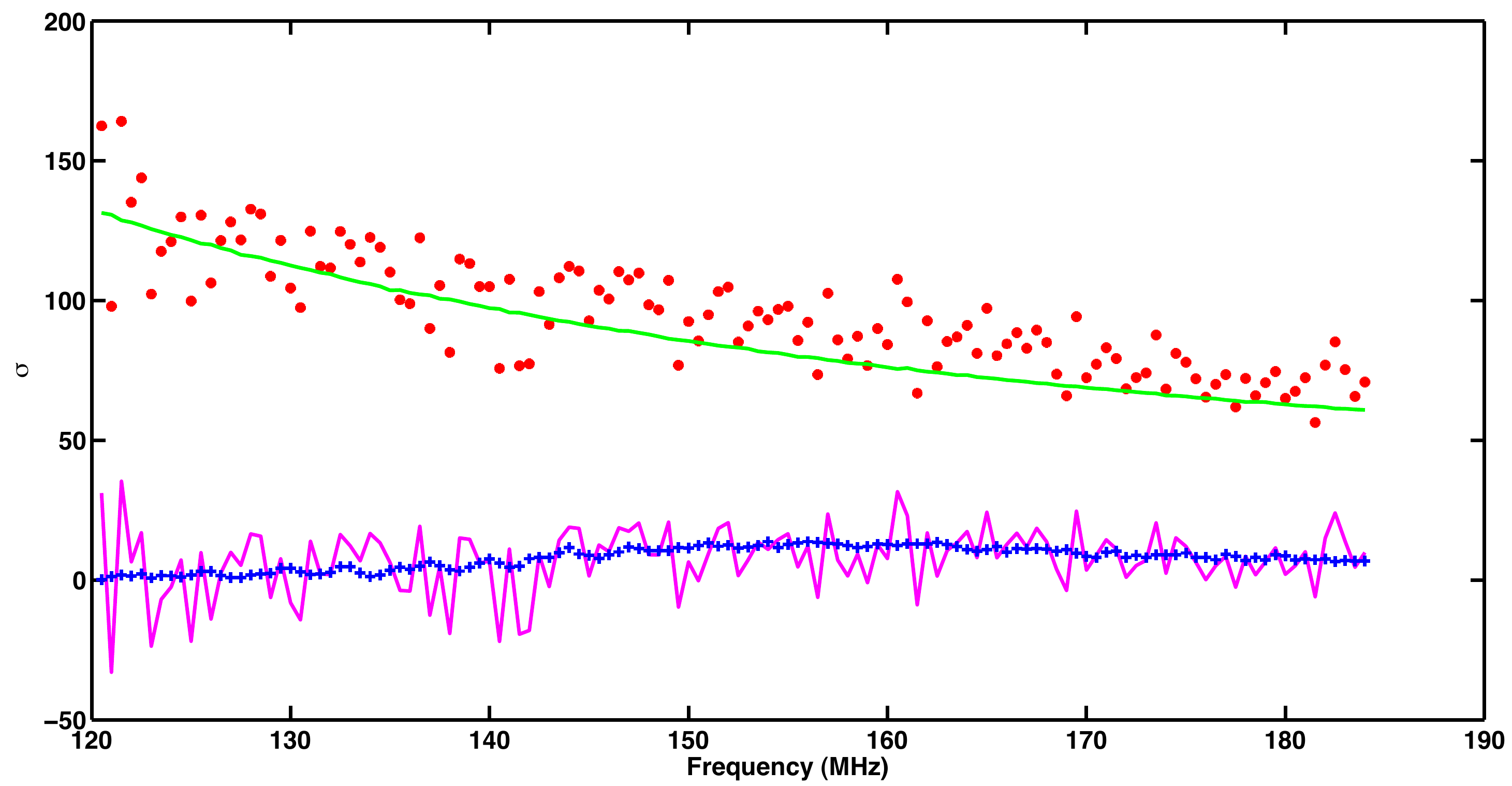}

\caption{Signal extraction using polynomials on the uv-plane. In the figure above the $rms$ as a function of frequency is shown. The red dots correspond to the measured  $rms$ of the residuals after foreground subtraction. The green line shows the $rms$ of the effective noise measured by differencing adjacent channels. The blue line is the evolution of the $rms$ of the cosmic signal and the magenta line the recovered signal.  }
\label{fig:ext}
\end{figure}

We achieve the best fit with a polynomial of 7th or 8th order which is higher than what \cite{harker09b}  and \cite{jelic08} suggest. This is due to the higher effective noise level that is present in the current simulations. In the aforementioned work the authors did not include the effects of calibration residual or the confusion noise and they assumed that the properties of the noise are known to a very high precision. In our case that does not hold true. The effective noise has different components that are affected in different ways by the data model and they also have different spatial, temporal and frequency behaviors. One way to quantify the noise properties is to use the difference of successive narrow-channel data. If the frequency resolution is high enough we can assume that the astrophysical signal does not change significantly and that the difference is purely determined by  the noise.  However, it is unfeasible to generate simulated data at such a high frequency resolution. To remedy this, we estimate the noise for each of the 128 channels of the simulation using the true underlying maps. We note that for the real observations, we will obtain very narrow frequency channels (10 kHz) from which we can accurately assess the noise level as function of frequency, time and baseline. We then estimate the PDF of the effective noise using bootstrapping. The result is shown in Figure \ref{fig:ext}.

\section{Conclusions}

To determine the theoretical limits and statistical efficiency  of the regularized maximum likelihood inversion, we used a Cram\'er-Rao bound analysis. We have shown  that the estimator becomes asymptotically efficient in the case of map-making with the LOFAR core and that imaging would require more than ten times better sensitivity.  This is within the grasp of the SKA. Based on our current simulations and understanding of the astrophysical and instrumental processes the LOFAR EoR experiment should be capable to detect the EoR signal, provided that the calibration can increase the dynamic range of the data with an error of less than $0.1$ per cent


\begin{thebibliography}{99}
 
\bibitem{furla} Furlanetto, S.~R., 
Oh, S.~P., Briggs, F.~H. \emph{Cosmology at low frequencies: The 21 cm 
transition and the high-redshift Universe}.\emph{Physics Reports}, {\bf{433}}, 181-301. 

\bibitem{panos} P.~Lampropoulos, \emph{The LOFAR Epoch of Reionization experiment data model},
PhD Thesis,  Groningen, 2010.

\bibitem{stoica} Stoica, ~P. and L.~R.~ Moses  \emph{Spectral Analysis of Signals},
Prentice Hall, 2005

\bibitem{stef} Wijnholds, ~S.  \emph{Fish-Eye Observing with Phased Array Radio Telescopes},
PhD Thesis, TU Delft, 2010

\bibitem{harker09b} Harker, G., and 14 
colleagues, \emph{Non-parametric foreground subtraction for 21-cm epoch of 
reionization experiments.}, \emph{Monthly Notices of the Royal Astronomical 
Society},  {\bf{397}}, 1138-1152. 


\bibitem{jelic08} Jeli{\'c}, V., and 12 
colleagues , \emph{Foreground simulations for the LOFAR-epoch of 
reionization experiment.}, \emph{Monthly Notices of the Royal Astronomical Society}, 
{\bf{389}}, 1319-1335. 


\end{thebibliography}
\end{document}